\documentclass[12pt]{iopart}
\usepackage{graphicx, caption}
\usepackage{multirow}
\usepackage{bm}
\usepackage{array}

\usepackage{hyperref}
\usepackage{lineno}
\usepackage{subfigure}
\usepackage{algorithm}
\usepackage{algorithmic}
\usepackage{amsbsy}
\usepackage{setspace}

\expandafter\let\csname equation*\endcsname\relax
\expandafter\let\csname endequation*\endcsname\relax
\usepackage {amsmath}

\usepackage[mathscr]{euscript}

\makeatletter

\newcommand{\Rmnum}[1]{\expandafter\@slowromancap\romannumeral #1@}
\makeatother

\usepackage{tablefootnote}
\usepackage{url}

\usepackage{amsfonts}
\usepackage{mathrsfs}

\usepackage{bbm}

\usepackage{color}
\usepackage{algorithm}
\usepackage{algorithmic}
\usepackage{graphicx}
\usepackage{epstopdf}
\usepackage{subfigure}
\usepackage{epsfig}
\usepackage{amssymb}
\usepackage{epsf}
\usepackage{subfigure}
\usepackage{multirow}
\usepackage{booktabs}
\usepackage{diagbox}

\begin{document}

\title[Dual-domain Attention-based Deep Network for Sparse-view CT Artifact Reduction]{Dual-domain Attention-based Deep Network for Sparse-view CT Artifact Reduction}

\author{Xiang~Gao}
\address{Research Center for Medical Artificial Intelligence, Shenzhen Institute of Advanced Technology, Chinese Academy of Sciences, Shenzhen, Guangdong 518055, China}
\address{University of Chinese Academy of Sciences, Beijing 101408,China}

\author{Ting~Su, Jiongtao~Zhu \& Jiecheng~Yang}
\address{Research Center for Medical Artificial Intelligence, Shenzhen Institute of Advanced Technology, Chinese Academy of Sciences, Shenzhen, Guangdong 518055, China}

\author{Yunxin~Zhang}
\address{Department of Vascular Surgery, Beijing Jishuitan Hospital, Beijing 100035, China}

\author{Donghua~Mi}
\address{Department of Vascular Neurology, Beijing Tiantan Hospital, Capital Medical University, Beijing 100070, China}

\author{Hairong~Zheng}
\address{Paul C Lauterbur Research Center for Biomedical Imaging, Shenzhen Institute of Advanced Technology, Chinese Academy of Sciences, Shenzhen 518055, China}

\author{ Xiaojing Long, Dong Liang \& Yongshuai Ge}
\address{Research Center for Medical Artificial Intelligence, Shenzhen Institute of Advanced Technology, Chinese Academy of Sciences, Shenzhen 518055, China}
\address{Paul C Lauterbur Research Center for Biomedical Imaging, Shenzhen Institute of Advanced Technology, Chinese Academy of Sciences, Shenzhen 518055, China}

\ead{ys.ge@siat.ac.cn}
\vspace{10pt}

\begin{abstract}
Due to the wide applications of X-ray computed tomography (CT) in medical imaging activities, radiation exposure has become a major concern for public health. Sparse-view CT is a promising approach to reduce the radiation dose by down-sampling the total number of acquired projections. However, the CT images reconstructed by this sparse-view imaging approach suffer from severe streaking artifacts and structural information loss. In this work, an end-to-end dual-domain attention-based deep network (DDANet) is proposed to solve such an ill-posed CT image reconstruction problem. The image-domain CT image and the projection-domain sinogram are put into the two parallel sub-networks of the DDANet to independently extract the distinct high-level feature maps. In addition, a specified attention module is introduced to fuse the aforementioned dual-domain feature maps to allow complementary optimizations of removing the streaking artifacts and mitigating the loss of structure. Numerical simulations, anthropomorphic thorax phantom and \textit{in vivo} pre-clinical experiments are conducted to verify the sparse-view CT imaging performance of the DDANet. Results demonstrate that this newly developed approach is able to robustly remove the streaking artifacts while maintaining the fine structures. As a result, the DDANet provides a promising solution in achieving high quality sparse-view CT imaging.
\end{abstract}

%
%
%
%
%

\section{INTRODUCTION}

Over the past half century, computed tomography (CT) has witnessed an irreplaceable role in modern medical imaging applications. However, concerns about the risks of excessive radiation exposure have attracted a lot of attention. According to the International Commission on Radiological Protection (ICRP), every increase of 1 mSv radiation dose in human body would rise the chance of canceration by nearly 1/20000\cite{1952American}. Nowadays, it has become a worldwide consensus to reduce the radiation dose of CT scans as low as reasonably achievable (ALARA). Dedicated researches from both the academic and industrial fields are performed to seek for the advanced low-dose CT imaging solutions .

Sparse-view CT scan is a promising approach to achieve low-dose CT imaging. By down-sampling the total number of the acquired projections, the radiation dose received by patients could be dramatically reduced. Different from the low-mAs scanning method\cite{2021AdaIN, 2018Aself,9468943}, usually, the X-ray tube current over the sparse-view CT scan is not reduced. Thus, the CT images reconstructed from such dataset would not suffer from significant quantum noise issues. Whereas, the sparse-view CT images reconstructed from the conventional filtered back-projection (FBP) algorithm would contain dramatic streaking artifacts and loss of anatomical structure due to the data incompletion. Therefore, improvement of the image quality in sparse-view CT imaging would be quite important for low-dose CT applications.

Over the past two decades, model-based iterative sparse-view CT image reconstruction methods were investigated by re-formulating the reconstruction task as a compression-aware optimization problem, in which certain prior information and mathematical optimization models are assumed to jointly remove the streaks. Because of the requirements of accurate forward model and parameter selections, the iterative CT reconstruction algorithms\cite{TV, 2012Low,2014Iterative} may have limitations in generating high-quality CT images. Additionally, they may also have the drawback of long running time in certain applications.

Recently, the prosperity of the deep learning (DL) techniques unveils many opportunities in medical image reconstruction fields, and have demonstrated excellent performance in low-dose CT reconstruction. In the sparse-view CT imaging, majority of the previous studies focus on removing the streaking artifacts in the CT image domain. For example, Jin et al.\cite{2016Deep} applied the U-net, and Han et al.\cite{2018Framing} proposed the dual frame U-net to perform sparse-view CT image reconstruction. Zhang et al.\cite{2018DDNet} proposed the DDNet to joint the advantages of Dense-net\cite{2016Densely} and deconvolution operation. Kang et al.\cite{2016A,2017Wavelet} combined the wavelet transform, residual block\cite{2016Resnet} and CNN together to remove the streaks. As a contrary, the sinogram based networks can also be utilized. For example, the sparse-view sinogram is interpolated into the full-view one via the U-net in SSNet\cite{2018SSNet}. Li et al.\cite{2019Learning} proposed the iCT-net to convert a sparse-view sinogram directly into a high-quality CT image. Lin et al.\cite{2019A} decomposed the sparse-view inverse problem into a groups of simple transformations, and used a layered network to perform CT reconstructions.

Despite of the promising performance, one apparent limitation of the above CNN approaches is that they barely process the image-domain and the sinogram-domain information at the same time. Instead, only the single-domain information is considered and processed. To mitigate such drawback, recently, the dual-domain networks that are able to share mutual information have attracted many research interests. As in the CT imaging field, for instance, Wu et. al.\cite{2021DRONE} integrated a dual-domain (CT image domain and sinogram domain) sparse-view CT reconstruction network with the iterative reconstruction procedure, and Lin et al.\cite{8953298} applied the dual-domain CNN to remove the metal artifacts.

Motivated by these aforementioned dual-domain studies, we herein propose a novel dual-domain attention deep network (DDANet) for high quality sparse-view CT reconstruction. In DDANet, a unique side-by-side network structure that ensures simultaneous feature extractions from the CT image and the sinogram is developed. Specifically, the U-Net is used to extract the features of CT images, the fully connected layer and atrous spatial pyramid pooling (ASPP)\cite{2018Encoder} module is used to extract the sinogram features, and the attention strategy is used to fuse the dual-domain features. The parallel feature extraction and high-level feature fusion allow information complement between the dual domains. The main contributions of this work are as follows:

\begin{itemize}
	\item A novel side-by-side dual-domain network with a paralleled architecture is proposed. It shows outstanding performance in removing the sparse-view CT image artifacts. 
	\item The attention mechanism is utilized to efficiently fuse the unique information learned from the sinogram domain and the CT image domain.
	\item The ASPP module with enlarged receptive field is employed to perform feature extractions in the sinogram domain. 	
\end{itemize}

\section{METHOD}
\subsection{Problem Formulation}
\begin{figure*}[h]
	\centering
	\includegraphics[width=6in]{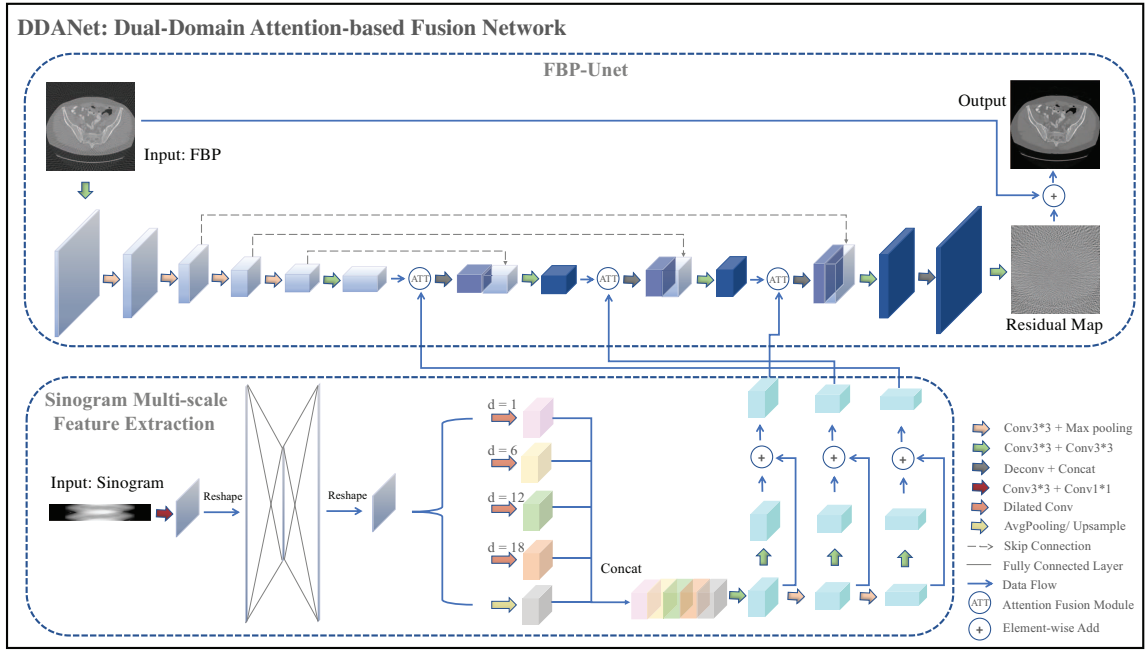}
	\caption{Architecture of the DDANet. Parameter $d$ represents the dilation rate of the dilated convolution layer. Note that only the FBP-Unet module and the sinogram multi-scale feature extraction module are depicted.} 
	\label{network}
\end{figure*}

In this sparse-view CT artifact removal problem, there are two network inputs: the CT image $\bm{F} \in {\mathbb{R}^{{{\rm{W}}_{img}} \times {{\rm{H}}_{img}} \times {{\rm{C}}_{img}}}}$ and the sinogram $\bm{S} \in {\mathbb{R}^{{{\rm{W}}_{proj}} \times {{\rm{H}}_{proj}} \times {{\rm{C}}_{proj}}}}$. The network is denoted as a function $g(\cdot)$. The goal of this work is to feed the $F$ and $S$ into $g(\cdot)$ to get the high-quality CT image $\bm{Q} \in {\mathbb{R}^{{{\rm{W}}_{img}} \times {{\rm{H}}_{img}} \times {{\rm{C}}_{img}}}}$ with mitigated streaking artifacts, namely:
\begin{equation}
\label{Q}
{\rm{Q}} = g(F,S) = \mbox{OP}_{decode}(\mbox{OP}_{fusion}(\mbox{OP}_F(F),\mbox{OP}_S(S)))
\end{equation}
where $\mbox{OP}_{F}$ represents the feature extraction operator of $F$, $\mbox{OP}_{S}$ represents the feature extraction operator of $S$,  $\mbox{OP}_{fusion}$ represents the feature fusion operator, and $\mbox{OP}_{decode}$ represents the feature decoding operator, see Fig.~\ref{network} and Fig.~\ref{attention}.

The overall structure of $g(\cdot)$ is shown in Fig.~\ref{network}. It is a side-by-side dual-domain input network. The high-level sinogram information is extracted independently to assist the artifact removal in the CT image domain. The network output is a residual image, which is added back to the input FBP image to mitigate the sparse-view streaking artifacts. In addition, the attention-based feature fusion module, i.e., $\mbox{OP}_{fusion}$, is illustrated in Fig.~\ref{attention}.

\subsection{FBP-Unet Module}
\label{startsample}
This component corresponds to the $\mbox{OP}_{F}$ operator in Eq.~(\ref{Q}). The feature extractions are performed via the U-net\cite{2015Uet}. In particular, feature maps with certain sizes (represented by $\{(w_1,h_1,c_1),(w_2,h_2,c_2),(w_3,h_3,c_3)\}$) of the CT images are obtained from each down-sampling stage. These features will be used for the subsequent feature fusions.

\section{Sinogram Multi-scale  Feature Extraction Module} 
This component corresponds to the $\mbox{OP}_{S}$ operator in Eq.~(\ref{Q}). By design, a fully connected layer is utilized to convert the sinogram onto the CT image domain. Different from the convolution, the fully connected layer is demonstrated to be more suitable for domain transformation \cite{2018Nature}. In addition, a layer with fewer nodes is added as an intermediate layer to reduce the total number of parameters. Dilated convolutions with different rates (d=1, d=6, d=12, d=18) are applied to enlarge the areas of the receptive fields. In fact, the receptive field of a dilated convolution with a kernel size $k_d$ and dilation rate d is equivalent to a convolution with a kernel size of $k_c$, namely,
\begin{equation}
k_c = k_d + (k_d - 1) * (d - 1)
\label{dilation}
\end{equation}
Moreover, average pooling is also implemented. By doing so, the network receptive field can be easily expanded without significantly increasing the total number of network parameters. Next, the dilated feature maps and the pooled feature maps are concatenated and fed into the multi-scale sinogram features extraction network module. Correspondingly, three features with sizes of $\in \{(w_1,h_1,c_1),(w_2,h_2,c_2),(w_3,h_3,c_3)\}$ are extracted. 

\subsection{Attention-based Feature Fusion Module}
\begin{figure*}[t]
	\centering
	\includegraphics[width=6in]{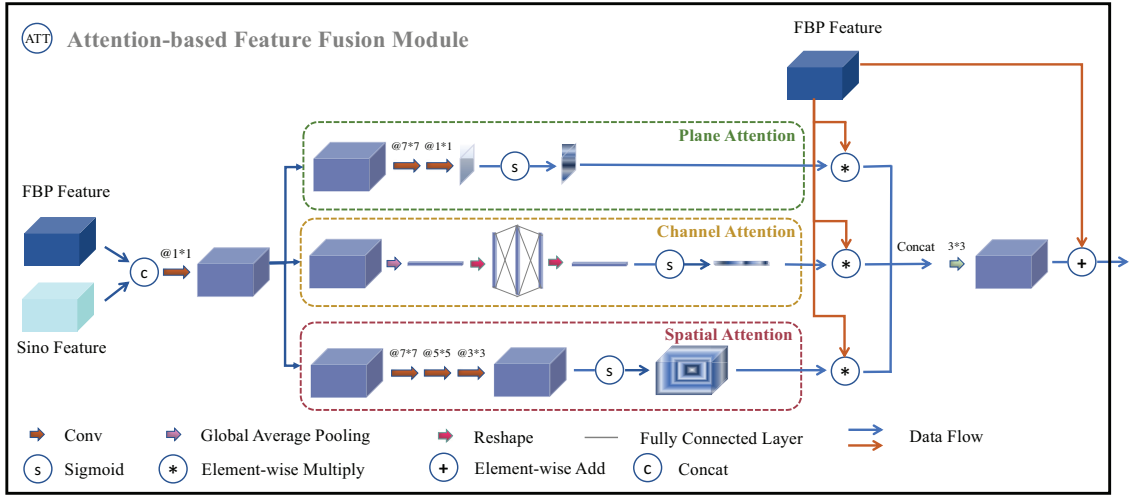}
	\caption{Structure of the attention-based feature fusion module. Three different attention mechanisms are used: plane attention, channel attention, and spatial attention.} 
	\label{attention}
\end{figure*}

This component corresponds to the $\mbox{OP}_{fusion}$ operator in Eq.~(\ref{Q}), and is denoted as $\mbox{ATT}$ in Fig.~\ref{network}. Specifically, a convolutional layer with a kernel size of 1 $\times$ 1 is used to extract the fused features $\bm{I_{fused}}$ from the CT image feature $\bm{I_{f}}$ and sinogram feature $\bm{I_{s}}$. The $\bm{I_{f}}$, $\bm{I_{s}}$ and $\bm{I_{fused}}$ belong to ${\mathbb{R}^{{{\rm{W_i}}} \times {{\rm{H_i}}} \times {{\rm{C_i}}}}}$, where ${{\rm{W}}_i},{H_i},{C_i} \in \{ w,h,c|({w_{1}},{h_{1}},{c_{1}}),({w_{2}},{h_{2}},{c_{2}}),({w_{3}},{h_{3}},{c_{3}})\}$. Next, the plane attention module $P_{att}$, channel attention module $D_{att}$, and spatial attention module $S_{att}$ are aggregated together to generate complete attention maps, which are multiplied with image-domain feature maps extracted from the FBP-Unet module, see more details in Fig.~\ref{attention}. As a result, certain spatial information that are beneficial for CT reconstruction are strengthened, and those irrelevant information are weakened.

$\noindent$ \textbf{Plane  Attention Module} extracts the plane feature map $\ M_p = [n_{1,1}, ...,n_{1,H_i},...,n_{W_i,1},...,,..., n_{W_i,H_i}] \in {\mathbb{R}^{{{\rm{W_i}}} \times {{\rm{H_i}}} \times {{\rm{1}}}}}$ via two convolutional layers. The sigmoid function is used to generate the final plane-attention map  $\bm \hat{M_p} = [\hat{n}_{1,1},...,\hat{n}_{1,H_i}, ..., \hat{n}_{W_i, 1}, ...,  \hat{n}_{W_i, H_i}] $.
\begin{equation}
{M_p} =  Conv_{7*7,1*1}({I_{fused}})), {\hat M_p} = \sigma ({M_p})
\label{P}
\end{equation}
where $Conv$ denotes the convolutional layer, and $\sigma$ denotes the sigmoid activation function. Afterwards, the $\hat{M_p}$ is applied to the input FBP feature $\bm {I_{f}} = [p_{1,1},...,p_{1,H_i}, ..., p_{W_i,1},..., p_{W_i,H_i}] $ with $ p \in {\mathbb{R}^{ {{\rm{C_i}}}}}$ using plane-wise multiplication to obtain a feature map with different plane weights.
\begin{equation}
{I_{f_{PA}}}= ATT_P({I_f},{\hat M_p})  = [{p_{1,1}}{\hat n_{1,1}},...,{p_{W_i,1}}{\hat
	n_{W_i,1}},...,\  {p_{1,H_i}}{\hat n_{1,H_i}}, ...,{p_{{{\rm{W}}_i},}}{\hat n_{{C_i}}}]
\label{P}
\end{equation}
$\noindent$ \textbf{Channel  Attention Module} squeezes the feature $\bm {I_{fused}} $  to obtain $M_c = [m_1, m_2, ..., m_{C_i}] \in {\mathbb{R}^{ {{\rm{C_i}}}}}$ via a global average pooling of dimension ${\mathbb{R}^{{{\rm{W_i}}} \times {{\rm{H_i}}} }}$ and two fully connected layers. The $M_c$ is activated by the sigmoid function to generate the channel-attention map $\bm \hat{M_c} = [\hat{m}_1, \hat{m}_2, ..., \hat{m}_{C_i}] $, namely,
\begin{equation}
\begin{aligned}
{M_c} = fc(A({I_{fused}})),    
{\hat M_c} = \sigma ({M_c})
\end{aligned}
\label{C}
\end{equation}
where $fc$ denotes two fully connected layers with total node numbers of $C_i / 2$ and $C_i$, A denotes the global average pooling, and $\sigma$ denotes the sigmoid activation function. The $\hat{M_c}$ is applied on the input CT feature $\bm {I_{f}} = [c_1, c_2, ..., c_{C_i}] $ with $c \in {\mathbb{R}^{{{\rm{W_i}}} \times {{\rm{H_i}}}}}$ using channel-wise multiplication to obtain a feature map with different channel weights.
\begin{equation}
\begin{aligned}
{I_{f_{CA}}} = ATT_C({I_f},{\hat M_c})= [{c_1}{\hat m_1}, {c_2}{\hat m_2},..., {c_{{C_i}}}{\hat m_{{C_i}}}]
\end{aligned}
\label{CA}
\end{equation}
$\noindent$ \textbf{Spatial  Attention Module} enhances the level of attention. The spatial feature map $\ M_s = [z_{1,1,1},..., z_{W_i,H_i,C_i}] \in {\mathbb{R}^{{{\rm{W_i}}} \times {{\rm{H_i}}} \times {{\rm{C_i}}}}}$ is obtained from certain network convolution. The sigmoid activation function is used to generate the $\hat{M}_s = [\hat z_{1,1,1},...,\hat z_{W_i,H_i,C_i}] \in {\mathbb{R}^{{{\rm{W_i}}} \times {{\rm{H_i}}} \times {{\rm{C_i}}}}}$ map:
\begin{equation}
\begin{aligned}
{M_s} =  Conv_{7*7,5*5,3*3}({I_{fused}})),    
{\hat M_s} = \sigma ({M_s})
\end{aligned}
\label{P}
\end{equation}
The $\hat{M_s}$ is applied on the input CT feature $\bm {I_{f}} = [s_{1,1,1}, ..., s_{W_i,H_i,C_i}] $ with $s \in {\mathbb{R}^1}$ using channel-wise multiplication to obtain a feature map with different channel weights.
\begin{equation}
\begin{aligned}
{I_{f_{SA}}} = ATT_S({I_f},{\hat M_s})= [{s_{1,1,1}}{\hat z_{1,1,1}},,..., {s_{{W_i,H_i,C_i}}}{\hat z_{{W_i,H_i,C_i}}}]
\end{aligned}
\label{CA}
\end{equation}

\begin{figure}[htbp]
	\centering
	\includegraphics[width=5in]{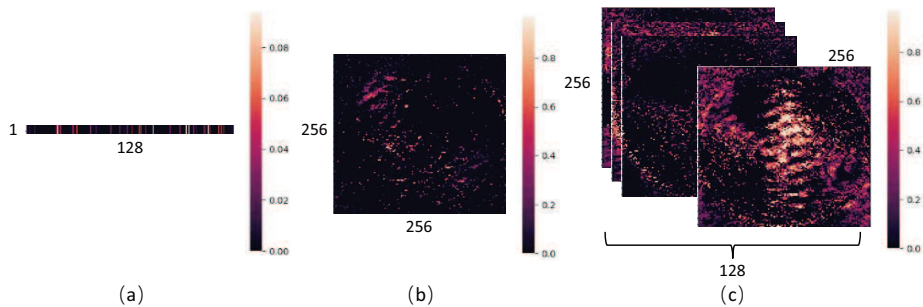}
	\caption{An example of the attention maps ($ \in {{\mathbb{R}^{{{\rm{256}}} \times {{\rm{256}}} \times {{\rm{128}}}}}}$) obtained from the proposed DDANet network: (a) the channel attention map, (b) the plane attention map, (c) the spatial attention map.} 
	\label{map_show}
\end{figure}

Finally, these unique attention features obtained from the above three different attention modules are superimposed along the depth dimension, and convolution is used to fuse the complex information. As shown in Fig.~\ref{map_show}, each attention map has a certain weight distribution.

\subsection{Network Training}

\subsubsection{Network Loss Function}
The network loss is a weighted summation of root-mean-square error (RMSE), structural similarity (SSIM) and peak signal-to-noise ratio (PSNR), see Eq.~(\ref{loss}). Herein, 
\begin{equation}
{L_{loss}} = {\lambda _1}{L_{rmse}} + \frac{{{\lambda _2}}}{{{L_{{\rm{ssim}}}}}
} + \frac{{{\lambda _3}}}{{{L_{psnr}}}}
\label{loss}
\end{equation}
where ${\lambda _1}$, ${\lambda _2}$, and ${\lambda _3}$ present the corresponding weight, RMSE is used to control the global difference, SSIM is used to control the structural information, and PSNR is used to control the degree of noise.

\subsubsection{Network Training Details}

The Adam optimizer was used, the initial learning rate was set to $1\times10^{-4}$ with decay rate of 0.95 every epoch. The network training were performed on a single Nvidia RTX A4000 GPU card.

\section{EXPERIMENTS}
\subsection{Dataset}
\subsubsection{Numerical data}
The data was prepared from the low dose CT images, which were published by the American Association of Physicists in Medicine (AAPM) low dose CT challenge\cite{AAPM}. The forward projections and CT image reconstructions were performed in Python with self-developed operators. Specifically, the distance from the X-ray source to the rotation center is 1000~mm, and is 1200~mm to the detector. There are 1024 detector elements with element of  0.6~mm$\times$0.6~mm. The  pixel size CT image is 0.625~mm $\times$ 0.625~mm. In total, 5410 training data and 526 testing data were generated. By default, 128 sparse-view projections were simulated.

\subsubsection{Experimental data}
Experimental data was acquired from an in-house CT imaging bench. The system was equipped with a rotating-anode X-ray tube (Varex G-242, Varex Imaging Corporation, UT, USA) and a flat panel detector (Varex 4343CB, Varex Image Corporation, UT, USA). The effective detector pixel size was $417 \mu m \times 417 \mu m$. The distance from the X-ray focal spot to the rotation center was 1156.3~mm, and was 1560.6~mm to the detector plane. The number of projections in the full-view CT scan was 900, with angular interval of 0.4 degree. The number of projections in the sparse-view CT scan was 128. An anthropomorphic thorax phantom (Model RS-111T, Radiology Support Devices, CA, USA) and an \textit{in vivo} anaesthetized monkey were scanned. The animal experiment was performed in compliance with the protocol (SIAT-IACUC-201228-YGS-LXJ-A1498, 5 January 2021) that was approved by the Institutional Animal Care and Use Committee (IACUC) of the Shenzhen Institute of Advanced Technology at the Chinese Academy of Sciences.

\subsubsection{Data enhancement}
\begin{figure}[h]
	\centering
	\includegraphics[width=5in]{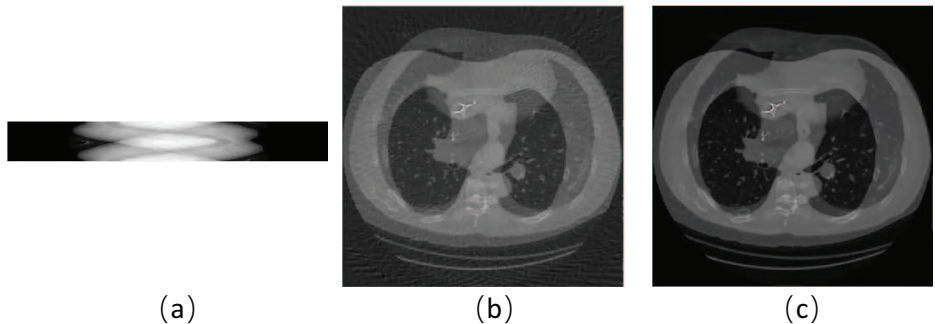}
	\caption{(a) Mixed sinogram with 128-view of projections, (b) mixed CT image reconstructed from (a), (c) ground truth of the mixed CT image in (b).} 
	\label{mixup_show}
\end{figure}

The mixup\cite{2017mixup} is used to augment the training data. Essentially, two individual images are mixed randomly over the network training with a certain weight:
\begin{equation}
\begin{aligned}
\label{mixup}
F_{mixed} = \lambda * F_{1} + (1-\lambda) * F_{2} \\
S_{mixed} = \lambda * S_{1} + (1-\lambda) * S_{2} \\
G_{mixed} = \lambda * G_{1} + (1-\lambda) * G_{2}
\end{aligned}
\end{equation}
where $\lambda$ denotes the weight coefficient sampled from the beta distribution. Symbols F, S and G denote the CT image, sinogram and the label, respectively. Every training data, has a 50\% chance of being mixed up with the other one.  One example of image mixup is shown in Figure \ref{mixup_show}. Be aware that the data mixup is not implemented on the test data.

\subsection{Ablation Experiments}
Ablation experiments were performed to prove the effectiveness of the core modules of DDANet. The quantitative results are shown in Table \ref{Tab01}. The baseline is the proposed DDANet without removing any module. In this case, RMSE is $4.2448 \times 10^{-4}$, SSIM is 0.9628, and PSNR is 42.7257.

\subsubsection{Dual Domain}
The dual-domain input strategy, in which the sinogram domain information is used to assist the streaking artifact removals in the CT image domain, is implemented. However, it is still unclear whether the dual-domain strategy is necessary. To address this question, the sinogram-domain inputs is replaced by the CT image domain features. According to the results in Table~\ref{Tab01}, it is easy to see that the dual-domain inputs indeed greatly improve the network performance. One possible reason is that the information loss in CT image domain can be compensated by the sinogram domain feature.

\subsubsection{Fully Connected Layer}
The fully connected layer in the sinogram feature extraction module was removed with the purpose to verify its effectiveness. All the other structures are kept unchanged. As seen in Table~\ref{Tab01}, the performance would be degraded compared to the DDANet. One possible explanation is that the sinogram and CT image contain different domain information, and the fully connected layer helps to perform the domain transformation and thus ease the fusion of different useful information, which are important for artifact removals.
\begin{table}[b]
	\scriptsize
	\centering
	\caption{Quantitative comparison results of the ablation experiments. The values of RMSE, PSNR and SSIM are estimated from the 526 test samples.}
	\label{Tab01}
	\begin{spacing}{1.5}
		\begin{tabular}{ccccc}
			
			\toprule
			\multirow{2}[3]{1.2cm}{\shortstack{Network}} &	\multirow{2}[3]{2.5cm}{\shortstack{Ablated Modules}} & \multicolumn{3}{c}{Performance}  \\ 
			\cmidrule(r){3-5}
			
			& & \multicolumn{1}{c}{RMSE }  & \multicolumn{1}{c}{SSIM }  &    \multicolumn{1}{c}{PSNR }   \\ 
			\midrule
			\multirow{5}{1.3cm}{DDANet\\(Ablation)} 
			& Dual-Domain & 4.5146$\times 10^{-4}$ &0.9598 &42.1833 \\
			& FC &4.3043$\times 10^{-4}$  & 0.9619 &42.6036   \\
			&Dilated Conv& 4.3349$\times 10^{-4}$  &0.9617 &42.5373  \\
			&Attention & 4.4249$\times 10^{-4}$ &0.9578 &42.3586  \\
			&Mixup & 4.3961$\times 10^{-4}$  & 0.9603 &42.4156  \\ 
			\cline{1-5}
			DDANet& -  & {\bfseries 4.2448$\times 10^{-4}$} &{\bfseries 0.9628} & {\bfseries 42.7257} \\
			\bottomrule
		\end{tabular}
	\end{spacing}
\end{table}
\subsubsection{Dilated Convolution}
The dilated convolution and the pooling operation (ASPP module) in the DDANet is removed to verify the effectiveness of the dilated convolution. As seen in Table \ref{Tab01}, the DDANet without dilated convolution shows worse performance. One possible reason is that the streaking artifacts are global ones, and need to be mitigated via global operations. The dilated convolution exactly better takes this task than the conventional pooling and convolution operations.

\subsubsection{Attention Fusion Module}
The output of DDANet is designed to be the streaking artifact only image, as a consequence, the attention fusion module plays the role of enhancing the extractions of streaking artifact feature information. As the results listed in the fourth and sixth rows of Table~\ref{Tab01}, clearly, the use of attention mechanism can boost the performance of DDANet in helping the intermediate layers be more focused on extracting the artifact related features.

\subsubsection{Mixup Data Augmentation}
Compared to the conventional data augmentation approaches such as image scaling and image rotation, the image mix-up can improve the model's capability in capturing the artifacts distributions, and enhance its robustness to different object structures. As can be observed from Table~\ref{Tab01}, the training data mix-up results in an improvement of $0.1513 \times 10^{-4}$, 0.0025 and 0.3101 in RMSE, SSIM, and PSNR, respectively.

\subsection{Comparison experiments}
The TV\cite{TV}, FBPConvNet\cite{2016Deep},  RedCNN\cite{2017REDCNN}, DDNet\cite{2018DDNet}, 
Frame Unet\cite{2018Framing}, SS-Net\cite{2018SSNet} methods are performed to compare with the DDANet. Essentially, these algorithms can be divided into three categories: the TV method belongs to the iterative algorithm, the FBPConvNet, RedCNN, DDNet and Frame Unet methods represent the image-domain only post-processing strategy, and the SS-Net method represents the sinogram-domain only post-processing strategy.

\section{RESULTS}

\begin{table}[b]
	\scriptsize
	\centering	
	\caption{Quantitative comparison results for the 128-view CT imaging. The averaged performance of the 526 testing images and the performance of the two selected cases in Fig.~\ref{Method_compare} are evaluated.}
	\label{Tab02}
	\begin{spacing}{1.5}
		\setlength\tabcolsep{2pt} 
		\begin{tabular}{ccccccccccc}
			\toprule
			\multirow{3}[3]{1.5cm}{\shortstack{Type}} &	\multirow{3}[3]{1cm}{\shortstack{Method}} & \multicolumn{3}{c}{Group Mean} & \multicolumn{3}{c}{Case \#1} & \multicolumn{3}{c}{Case \#2} \\
			\cmidrule(r){3-5} \cmidrule(r){6-8} \cmidrule{9-11}
			& &\multirow{2}{0.75cm}{RMSE \\$(10^{-4})$} & \multirow{2}{0.75cm}{SSIM }  &  \multirow{2}{0.75cm}{PSNR }   & \multirow{2}{0.75cm}{RMSE \\$(10^{-4})$} &   \multirow{2}{0.75cm}{SSIM }   &   \multirow{2}{0.75cm}{PSNR } & \multirow{2}{0.75cm}{RMSE \\$(10^{-4})$} &   \multirow{2}{0.75cm}{SSIM }  &   \multirow{2}{0.75cm}{PSNR } \\ \\
			\midrule
			
			\multirow{1}{2.2cm}{Analytical} & FBP &15.374&0.5836&31.5229&11.412&0.6094&31.5333&12.762&0.5250&29.5876 \\
			\cline{1-11}
			\multirow{1}{1.8cm}{Iterative} & TV\cite{TV} &4.6465&0.9541&41.9347&3.5378&0.9423&40.6055&3.2527&0.9429&40.1288 \\
			\cline{2-11}
			\multirow{4}{2.4cm}{Image-Domain}
			&FBPConvNet\cite{2016Deep} & 4.7517  & 0.9544& 41.7357&3.5552&0.9402&40.1832&3.0470&0.9489&40.1500 \\
			&RedCNN\cite{2017REDCNN}&5.1121 & 0.9459& 41.0911 &3.6973&0.9327&39.6745&3.2117&0.9388&39.4363\\
			&DDNet\cite{2018DDNet}& 4.7141& 0.9545&41.8076 &3.5760&0.9402&40.2092&3.0476&0.9493&40.1932 \\
			&Frame Unet\cite{2018Framing} &4.6920 & 0.9554 &41.8470 &3.6032&0.9411&40.2065&3.1123&0.9494&40.2103 \\ 
			\cline{2-11}
			\multirow{1}{2.7cm}{Sinogram-Domain}
			&SS-Net\cite{2018SSNet} & 7.6899 & 0.9033 & 37.5562 &5.3517 &0.8819 & 36.4759 & 4.7687 & 0.8945 & 36.4387\\ 
			\cline{1-11}
			\multirow{1}{2.3cm}{Dual-Domain}	& DDANet  & {\bfseries 4.2448} &{\bfseries 0.9628} & {\bfseries 42.7257} &{\bfseries 3.2861}&{\bfseries 0.9482}&{\bfseries 40.8635}&{\bfseries 2.7069}&{\bfseries 0.9596}&{\bfseries 41.2300}\\
			\bottomrule
			
		\end{tabular}
	\end{spacing}
\end{table}

\subsection{Numerical results}
\begin{figure*}[t]
	\centering{}
	\includegraphics[width=5in]{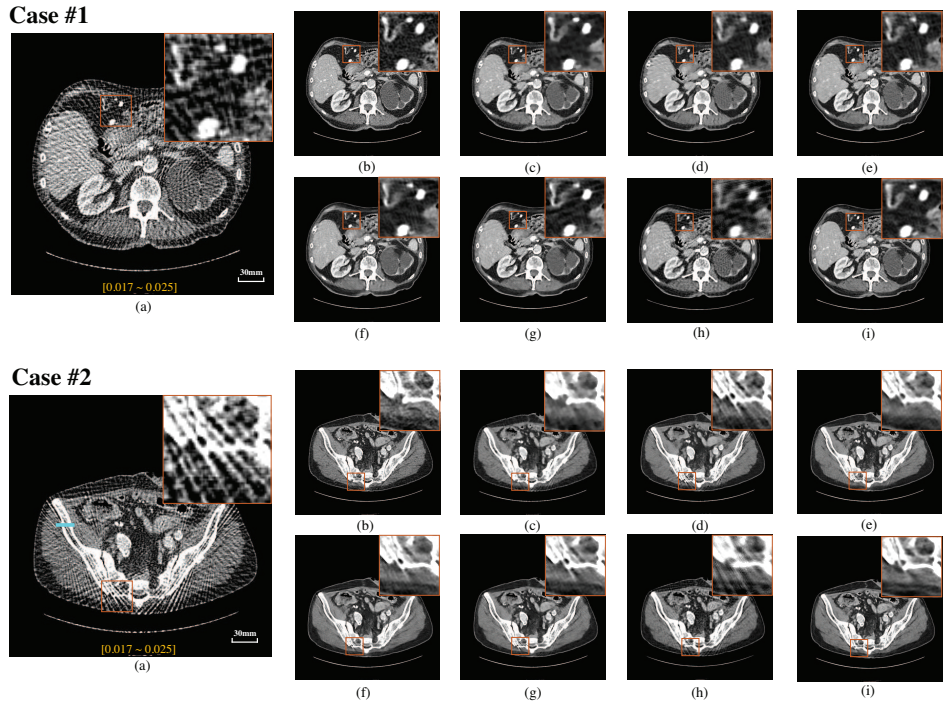}
	\caption{Results of simulated CT images for different reconstruction methods: (a) FBP, (b) Ground truth, (c) TV, (d) RedCNN, (e) FBPConvNet, (f) Frame Unet, (g) DDNet, (h) SS-Net, (i) DDANet. The display window is [0.017, 0.025] mm$^{-1}$.  The scale-bar denotes 30 mm.} 
	\label{Method_compare}
\end{figure*}

\begin{figure}[h]
	\centering{}
	\includegraphics[width=4.7in]{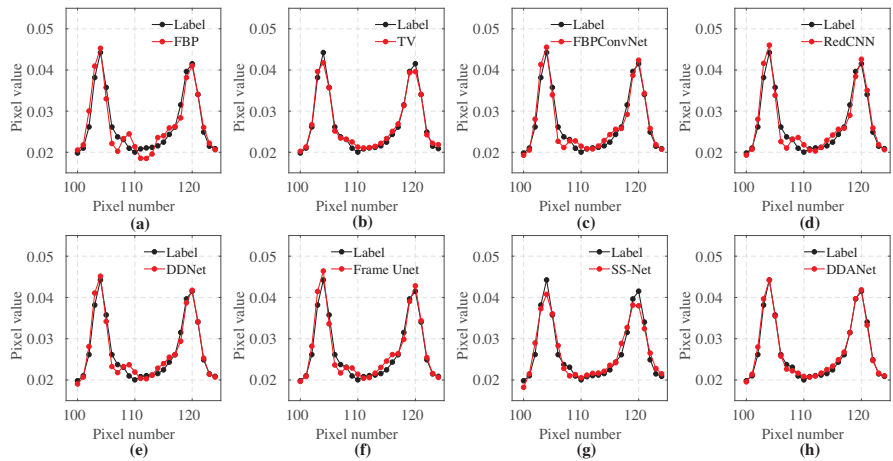}
	\caption{The profiles along the horizontal green line in Fig.~\ref{Method_compare}(a) (Case \#2). The red curves denote the ground truth, and the blue curves denote the different methods: (a) FBP, (b) TV, (c) RedCNN, (d) FBPConvNet, (e)Frame Unet, (f) DDNet, (g) SS-Net, (h) DDANet.}
	\label{profile}
\end{figure}

The numerical performance of DDANet is validated on the AAPM low dose CT data, and results are presented in Fig.~\ref{Method_compare}. As seen, the TV algorithm removes most of the streaking artifacts, but blurs the fine structures. The single-domain methods (including image-domain method and sinogram-domain method) have difficulty in removing substantial streaks while preserving fine details of the image. As a contrary, most of the streaks can be nicely removed by the developed DDANet without sacrificing the image sharpness. In addition, line profiles of the high-contrast bone (highlighted in blue in Fig.~\ref{Method_compare}(a), Case \#2) were also compared, see Fig.~\ref{profile}. It is observed that the DDANet method generates the most consistent profile comparing to the ground truth, indicating the best capability of maintaining the signal precision and image sharpness. The quantitative analysis results are listed in Table \ref{Tab02}. Again, the DDANet achieves the smallest RMSE value, and the highest SSIM and PSNR values.

\subsection{Results of robustness study}
\begin{figure*}[!h]
	\centering{}
	\includegraphics[width=6in]{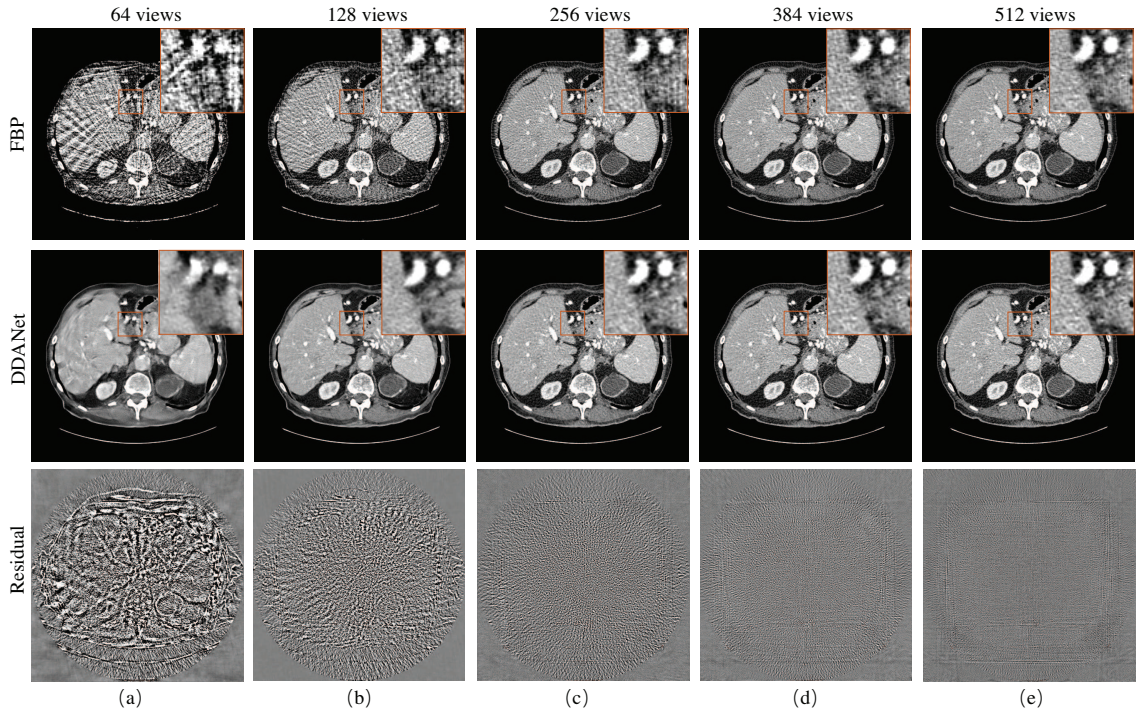}
	\caption{Results of different number of projections. Images in the first row are reconstructed by the FBP algorithm, and images in the second row are reconstructed by the DDANet. The residual images of the ground truth and DDANet are depicted in the third row. The display window is [0.017, 0.025] mm$^{-1}$ for the CT images, and [-0.001, 0.001] mm$^{-1}$ for the residual images.} 
	\label{views}
\end{figure*}

The robustness of the proposed DDANet network in removing the streaking artifacts over different sparse-view projection data, i.e., 64, 256, 384 and 512, are investigated, see the results in Fig.~\ref{views}. As observed, the image quality gets less satisfactory when the number of projections is 64. However, DDANet is still able to remove most of the streaking artifacts and keep the majority of structural information intact under this very challenging condition. As the number of projections increases, the quality of the reconstructed CT images from DDANet boosts gradually, and the residual losses are found to be negligible. Additionally, the quantitative analysis results are listed in Table~\ref{Tab03}. Be aware that certain convolutions and pooling layers are altered in the sinogram feature extraction module for those sinograms with different dimensions (not equal to 128) to keep consistent feature sizes.

\begin{table}[htb]
	\scriptsize
	\centering
	\caption{Quantitative comparison results for different views (68, 128, 256, 384, 512).}
	\label{Tab03}
	\begin{spacing}{1.5}
		\begin{tabular}{cccc}
			\toprule
			\multirow{2}[3]{0.8cm}{\shortstack{Views}}  & \multicolumn{2}{c}{FBP / DDANet}  \\
			\cmidrule(r){2-4}
			&  \multicolumn{1}{c}{RMSE $(10^{-4})$} &   \multicolumn{1}{c}{SSIM } &   \multicolumn{1}{c}{PSNR }    \\ 
			\midrule
			64 &27.801 / 6.7533 &0.3741 / 0.9297&26.3721 / 38.6759 \\
			128 &15.374 / 4.2448  &0.5836 / 0.9628&31.5229 / 42.7257 \\
			256 &7.9227 / 3.1028 &0.8343 / 0.9762 &37.2751 / 45.4483 \\
			384 &4.6774 / 2.3546&0.9338 / 0.9857 &41.8533 / 47.8435 \\
			512 &3.6605 / 1.8875 &0.9590 / 0.9893 &43.9816 / 49.7658 \\ 
			\toprule	
		\end{tabular}
	\end{spacing}
\end{table}

\subsection{Experimental results}
\begin{figure*}[!htbp]
	\centering{}
	\includegraphics[width=6in]{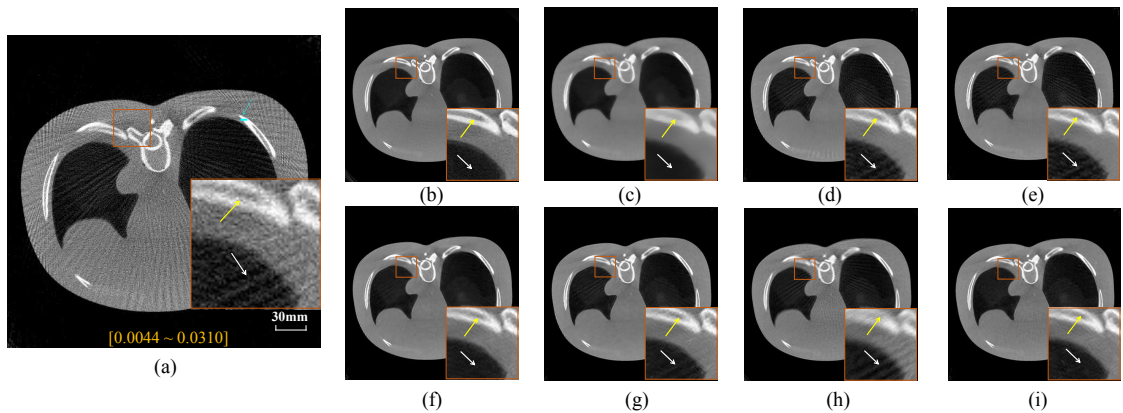}
	\caption{Experimental results of the thorax phantom: (a) FBP, (b) Ground truth, (c) TV, (d) RedCNN, (e) FBPConvNet, (f) FrameUnet, (g) DDNet, (h) SS-Net, (i) DDANet. The display window is [0.004, 0.031] mm$^{-1}$. The scale-bar denotes 30 mm.} 
	\label{Chest}
\end{figure*}

\begin{figure*}[!htbp]
	\centering{}
	\includegraphics[width=6in]{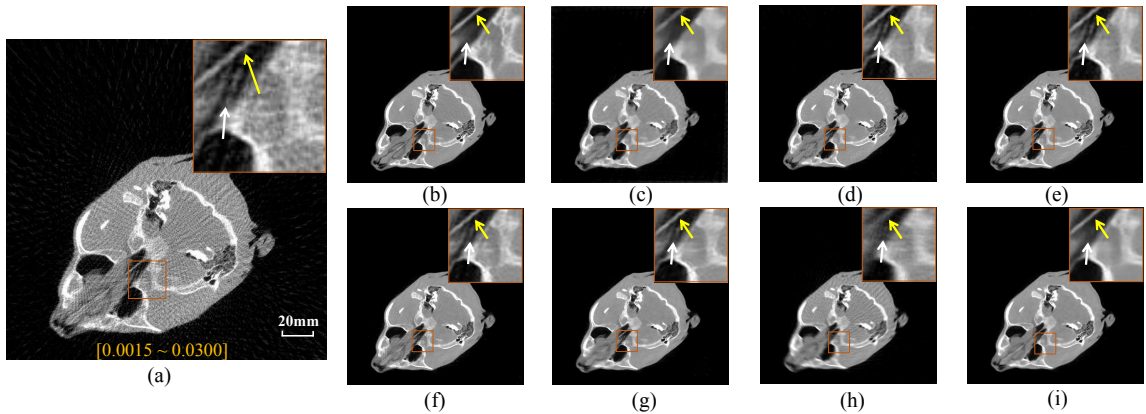}
	\caption{Experimental results of the monkey head:  (a) FBP, (b) Ground truth, (c) TV, (d) RedCNN, (e) FBPConvNet, (f) FrameUnet, (g) DDNet, (h) SS-Net, (i) DDANet. The display window is [0.001, 0.030] mm$^{-1}$. The scale-bar denotes 20 mm.} 
	\label{Monkey}
\end{figure*}

Two experimental sparse-view CT imaging data, an anthropomorphic thorax phantom and an anaesthetized monkey, are verified, see the results in Fig.~\ref{Chest} and Fig.~\ref{Monkey}. With regard to the selected region-of-interest (ROI) in Fig.~\ref{Chest}, it can be seen that the proposed DDANet outperforms the other methods in removing the streaking artifacts. Although the TV and FrameUnet methods are able to reduce more artifacts, the reconstructed images become quite blurry. For the animal experiments, again, the DDANet shows the best capability in eliminating the streaking artifacts, see the regions highlighted by the white arrows in Fig.~\ref{Monkey}. 

\begin{figure}[htbp]
	\centering{}
	\includegraphics[width=6in]{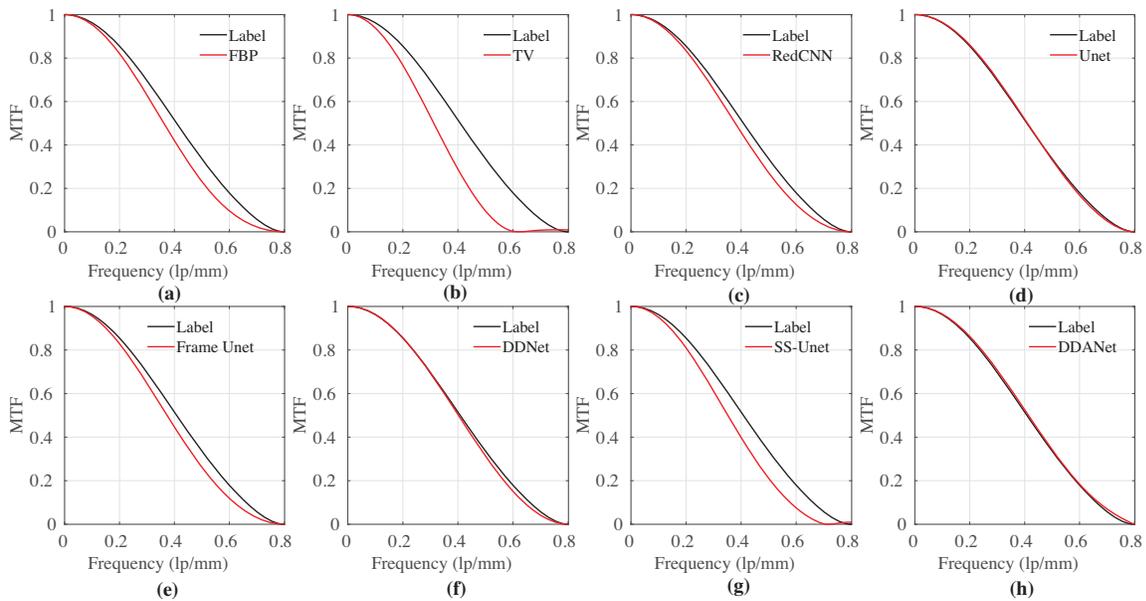}
	\caption{The MTF results. (a)FBP, (b) TV, (c) RedCNN, (d) FBPConvNet, (e)Frame Unet, (f) DDNet, (g) SS-Net, (h) DDANet.} 
	\label{MTF}
\end{figure}

Moreover, the image spatial resolution, i.e., the modulation transfer function (MTF), is compared quantitatively, see the plots in Fig.~\ref{MTF}. Specifically, the MTF curves are calculated from the highlighted bone/tissue region (cyan line) in Fig.~\ref{Chest}. Overall, the MTF curves further confirm the visual performance in Fig.~\ref{Chest} and Fig.~\ref{Monkey}. For example, the MTF curve generated from the TV algorithm gets narrower, indicating a certain loss of the image resolution. Compared with the reference FBP method (for full-view reconstruction), the DDANet method slightly degrades the image resolution. Note that the MTF is measured on the high-contrast object, thus the actual visual perforamance may get varied for low-contrast objects.

\section{DISCUSSION AND CONCLUSION}
In this work, a dual-domain attention network (DDANet) is proposed for the sparse-view CT imaging. Different from the reconstruction methods developed based on the single-domain information, DDANet is able to simultaneously utilize the dual-domain (CT image domain and sinogram domain) information. With such a particular network design, additional mutual information could be accessed to assist the sparse-view CT image reconstructions. To generate the best imaging performance, certain techniques are used to enhance the feature extractions. For instance, the fully connected layer is adopted to perform domain transformation between the sinogram and the CT image, the dilated convolutional layer is utilized to enlarge the receptive field, multiple attention mechanisms are incorporated to make feature fusions. The superior performance of the DDANet network are verified via additional ablation experiments. Finally, numerical simulations, phantom and animal experiments are conducted. Results demonstrate the advancement of the dual-domain based DDANet in reducing the streaking artifacts for sparse-view CT imaging.

The trained DDANet by numerical data is directly used on the experimental data without any network fine tuning. Due to this reason, the reconstructed experimental CT images may show inferior performance than the numerical simulations. Enhanced performance of the DDANet network would be expected if further network fine tuning could be implemented on real experimental data.


In future, the following topics can be investigated: (1) the self-attention mechanism defined in the transformer network\cite{2017transformer} could be used to carry out the global feature extractions, particularly for the sinogram-domain sub-network. By doing so, the removal of the streaking artifacts that spread over the entire CT image could become more effective. (2) the hint learning\cite{2015HintLearning} approach would be tested and incorporated to replace the fully connected layer in the sinogram-domain sub-network with the purpose of greatly shrinking the total size of the network parameters while minimally degrading the entire network performance.

In conclusion, a dedicated end-to-end dual-domain attention-based deep network (DDANet) is proposed to reconstruct high quality sparse-view CT images. The performance of DDANet is validated on the numerical simulations, anthropomorphic thorax phantom and \textit{in vivo} preclinical experiments. Results demonstrate that this newly developed network is able to remove the streaking artifacts robustly while maintaining the fine structural details in sparse-view low-dose CT imaging.


%

\section*{ACKNOWLEDGMENT}
The authors would like to thank Dr. Yaoqin Xie for borrowing the anthropomorphic thorax phantom. This work is supported by Guangdong Basic and Applied Basic Research Foundation (2019A1515011262, 2020A1515110685), Shenzhen Basic Research Program (Grant No.~JCYJ20200109115212546), Guangdong Key Area Research Program (2020B1111130001), National Natural Science Foundation of China (12027812, 11804356), Youth Innovation Promotion Association of Chinese Academy of Sciences (2021362).

\section*{References}
\bibliographystyle{unsrt}
\bibliography{IEEEexample}

\end{document}